\documentclass[onecolumn,showpacs]{revtex4}

\usepackage{graphicx}
\usepackage{dcolumn}
\usepackage{bm}

\input epsf

\begin{document}

\title{\Large A New Result on the Dynamical Symmetry of Spherical Collapse}

\author{\bf Subenoy
Chakraborty$^1$\footnote{subenoyc@yahoo.co.in}, Asit
Banerjee$^2$\footnote{asitb@cal3.vsnl.net.in} and Ujjal
Debnath$^3$\footnote{ujjaldebnath@yahoo.com}}

\affiliation{$^1$Department of Mathematics, Jadavpur University,
Calcutta-32, India.\\$^2$Department of Physics, Jadavpur
University, Calcutta-32, India.\\ $^3$Department of Mathematics,
Bengal Engineering and Science University, Shibpur, Howrah-711
103, India.\\}

\date{\today}

\begin{abstract}
A dynamical symmetry for spherical collapse has been studied using
a linear transformation of the initial data set (mass and kinetic
energy function) and the area radius. With proper choice of the
initial area radius, the evolution as well as the physical
parameters namely energy density and shear remain invariant both
initially and at any time instant. Finally, it is found that the
final outcome of collapse depends on the initial choice of the
area radius.
\end{abstract}

\pacs{04.20 Dw, 04.20 Ex}

\maketitle

Debnath, Chakraborty and Dadhich in a recent paper [1] while
demonstrating the true invariance of the collapse dynamics, chose
to scale the comoving radial coordinate also along with other
initial data sets. This has, in fact, destroyed the invariance of
energy density and shear as well as the collapse dynamics itself.
It can be shown clearly by the following arguments. In their case
the linear transformations are chosen in the form
$$
F\rightarrow a^{n-1}F ,~~f\rightarrow a^{2}f,~~ R\rightarrow
a~R~~\text{and}~~r\rightarrow a~r,
$$

where $a$ is a constant, $n$ is the total number of dimensions and
$r$ is the comoving radial coordinate. $F,~ f,~ R$ have the usual
meanings [2,3]. The new radial coordinate in reality changes the
magnitudes of the functions $F,~f$ and $R$. So when the above
scalings are performed the physical quantities such as
$\rho,~\sigma$ etc although retain their structural forms
unchanged in terms of $F,~f,~r$ etc they do suffer changes in
their magnitudes and can no longer be claimed invariant. The
scaling of the coordinate $r$ was, however, demanded by their
initial choice $R(t_{i},r)=r$. One can avoid such unwanted scaling
of $r$ if one chooses $R(t_{i},r)$ as another suitable function of
$r$. In the following text we have done exactly the same. So our
choice gives the true invariance of $\rho,~\sigma$ and the
dynamical equations for a class of data sets which lead to the
same end results as well as the evolution of the system chosen. We
should point out here that though our model is a spherical
Tolman-Bondi-Lema\^{\i}tre (TBL) model yet similar results may
also be valid in the quasi-spherical collapse as well.\\

The inhomogeneous spherically symmetric dust space-time of $n$
dimension can be represented by the following metric [4]
\begin{equation}
ds^{2}=dt^{2}-\frac{R'^{2}}{1+f(r)}dr^{2}-R^{2}d\Omega^{2}_{n-2}
\end{equation}

Here we have used comoving coordinates. In the above metric
$g_{00}$ (the co-efficients of $dt^{2}$)=1, because the dust
particle follows geodesics. The function $f(r)$ is related with
the curvature of the $(n-1)$ dimensional space part and fixes it
as bound, unbound and marginally bound for $f(r)<0$, $f(r)>0$ and
$f(r)=0$ respectively. The function $R(t,r)$ stands for the area
radius of the corresponding shell. The evolution equation is given
by
\begin{equation}
\dot{R}^{2}=f(r)+\frac{F(r)}{R^{n-3}}
\end{equation}

where $F(r)$ can be interpreted as the mass contained within the
comoving radius `$r$' and is reflected in the expression for
energy density $\rho(t,r)$:
\begin{equation}
\rho(t,r) =\frac{(n-2)}{2}~\frac{F^{\prime}}{R^{n-2}R^{\prime }}
\end{equation}

The shear scalar $\sigma(t,r)$ is given by [3,4]
\begin{equation}
\sigma=\sqrt{\frac{n-2}{2(n-1)}}~\frac{(R\dot{R}'-\dot{R}R')}{RR'}
\end{equation}

The above collapsing models are characterized by the initial data
given on the initial hypersurface $t=t_{i}$.  It is to be noted
that one can make an arbitrary relabeling of spherical dust shells
by $r\rightarrow \psi(r)$, without any loss of generality. We make
here the suitable choice of the radial coordinate such that at
$t=t_{i}$,
\begin{equation}
R(t_{i},r)=\frac{rF^{a}}{f^{b}}
\end{equation}

where, $a$ and $b$ are constants.\\

We now proceed to find out a class of initial data functions which
under linear transformation yield exactly the same density, shear
and evolution equation not only at the initial instant but also at
any arbitrary instant. These are [1,5]
\begin{equation}
\{F, ~f\}\rightarrow \{mF,~lf\}
\end{equation}

In order that evolution equation (2) remains invariant under the
above transformations we must require a relation between $m$ and
$l$ in the form
\begin{equation}
m=l^{\frac{n-1}{2}}
\end{equation}

and a scaling of the area radius as
\begin{equation}
R\rightarrow l^{1/2}R
\end{equation}

So finally we find that the transformations (6) and (8) in view
of (7) retain $\rho$ and $\sigma$ invariant. The interesting
point is that the above transformations do not alter the scaling
used for the radial coordinate given in (5) provided the
constants $a$ and $b$ are related as follows
\begin{equation}
a(n-1) -2b=1
\end{equation}

It is now expected that the collapse of the dust sphere would
start from a regular initial hypersurface $t=t_{i}$ and hence
$\rho_{i}$ and $\sigma_{i}$ should be regular on the initial
hypersurface and in particular at $r=0$.\\

The initial density and shear can now be given using (5). These
are
\begin{equation}
\rho_{i}=\rho(t_{i},r)=\frac{(n-2)}{2}~\frac{rF'}{F}\left[
r^{n-1}\frac{F^{(n-1)a-1}}{f^{(n-1)b}}\left\{1+ar~\frac{F'}{F}-br~\frac{f'}{f}
\right\} \right]^{-1}
\end{equation}
and
\begin{equation}
\sigma_{i}=\sigma(t_{i},r)=\frac{\sqrt{\frac{(n-2)}{8(n-1)}}
\left[\left\{\frac{rF'}{F}-(n-1)\left(
1+ar~\frac{F'}{F}-br~\frac{f'}{f}\right) \right\} +
r^{n-3}\frac{F^{(n-3)a-1}}{f^{(n-3)b-1}}\left\{\frac{rf'}{f}-2
\left( 1+ar~\frac{F'}{F}-br~\frac{f'}{f} \right) \right\} \right]}
{r^{\frac{n-1}{2}}\frac{F^{\frac{(n-1)a-1}{2}}}{f^{\frac{(n-1)b}{2}}}
\left(1+ar~\frac{F'}{F}-br~\frac{f'}{f}
\right)\left(1+r^{n-3}\frac{F^{(n-3)a-1}}{f^{(n-3)b-1}}
\right)^{1/2} }
\end{equation}

Now if we choose a power series function for $F(r)$ and $f(r)$
around $r=0$, we may conclude that both $\frac{rF'}{F}$ and
$\frac{rf'}{f}$ are finite at the centre and hence in order that
$\rho(t_{i},r)$ is regular at $r\rightarrow 0$, we must have

$$\begin{array}{c}
lim~~~r^{n-1}\frac{F^{(n-1)a-1}}{f^{(n-1)b}}=\text{a~
non-zero~finite~number.}\\
\hspace{-7cm}~r\rightarrow 0\\\\
\end{array}
$$

The regularity of the initial shear at $r\rightarrow 0$ further
demands in view of (11) that the quantity
$$\begin{array}{c}
lim~~~r^{n-3}\frac{F^{(n-3)a-1}}{f^{(n-3)b-1}}~~~~~\text{to~
be~zero~or~a~~finite~constant.}\\
\hspace{-8cm}~r\rightarrow 0\\\\
\end{array}
$$

Since our study is restricted to the region close to $r=0$ for
smooth initial data we choose the power series functions for
$F(r)$ and $f(r)$ in this region, so that we can write [4]
\begin{equation}
F(r)=\sum_{j=0}^{\infty}F_{j}~r^{\alpha+j}
\end{equation}
\begin{equation}
f(r)=\sum_{j=0}^{\infty}f_{j}~r^{\beta+j}
\end{equation}

Using the expressions (12) and (13) close to the centre $r=0$,
the condition worked out previously for the regularity of the
initial density demands $\alpha$ and $\beta$ to satisfy the
following relation
\begin{equation}
a\alpha-b\beta+1=\frac{\alpha}{n-1}
\end{equation}

We now proceed to examine the expression for shear given in (11).
In view of (12) and (13) and also (14) the quantity
$\frac{rF'}{F}-(n-1)\left(1+ar~\frac{F'}{F}-br~\frac{f'}{f}
\right)$ reduces to zero. So in order that $\sigma(t_{i},r)$
vanishes at the centre ($r=0$),
$r^{n-3}\frac{F^{(n-3)a-1}}{f^{(n-3)b-1}}$ must vanish at
$r\rightarrow 0$.\\

The above in turn leads to the condition
$$
(n-3)(1+a\alpha-b\beta)>\alpha-\beta,
$$

which when combined with the condition (14) imposes the following
restriction on the magnitudes of $\alpha$ and $\beta$ namely,
\begin{equation}
\frac{\alpha}{n-1}>\frac{\alpha-\beta}{n-3}
\end{equation}

provided we assume $\alpha>\beta$ and $n>3$.\\

So finally, we get the restriction in the form
\begin{equation}
\alpha>\beta>\frac{2\alpha}{n-1}
\end{equation}

We now proceed to write the expression for the initial energy
density near $r=0$ using the power series expansion forms (12) and
(13). It is given by
\begin{equation}
\rho_{i}=\rho(t_{i},r)=\frac{(n-1)(n-2)}{2}\frac{f_{0}^{b(n-1)}}{F_{0}^{2b}}
\left[1+\frac{b(\alpha+1)}{\alpha}\left\{(n-1)\frac{f_{1}}{f_{0}}-\frac{2F_{1}}{F_{0}}
\right\}~r+\rho_{2}~r^{2}+O(r^{3}) \right]
\end{equation}

where $\rho_{2}$ is actually a constant, the exact expression of
which is given later.\\

The equation (17) shows that for a smooth physically acceptable
behaviour of the density function at $r=0$ one can demand
$\left(\frac{\partial \rho_{i}}{\partial r} \right)_{r=0}=0 $ and
hence it follows that the constant
\begin{equation}
(n-1)\frac{f_{1}}{f_{0}}-\frac{2F_{1}}{F_{0}}=0
\end{equation}

The constant $\rho_{2}$ in (17) has the simplified expression
using (18) in the form
\begin{equation}
\rho_{2}=b\left(1+\frac{2}{\alpha} \right)\left\{
(n-1)\frac{f_{2}}{f_{0}}-\frac{2F_{2}}{F_{0}}
\right\}+2b\frac{F_{1}^{2}}{F_{0}^{2}}\left[
1+\frac{1}{\alpha}+\frac{\{2-2\alpha+b(n-1)\alpha \}}{(n-1)\alpha}
\right]
\end{equation}

Again if one assumes that $\rho_{i}$ is maximum at $r=0$ and
decreases with increasing $r$ one must have $\rho'_{i}<0$ in the
region close to $r=0$, which in turn demand $\rho_{2}<0$ near the
centre.\\

Now the solution of the evolution eq(2) can be written as [4]
\begin{equation}
t-t_{i}=\frac{2}{(n-1)\sqrt{F}}\left[
r^{\frac{n-1}{2}}\frac{F^{\frac{a(n-1)}{2}}}{f^{\frac{b(n-1)}{2}}}~_{2}F_{1}
[\frac{1}{2},d,d+1,-r^{n-3}\frac{F^{a(n-3)-1}}{f^{b(n-3)-1}}]
-R^{\frac{n-1}{2}}~_{2}F_{1}[\frac{1}{2},d,d+1,-\frac{fR^{n-3}}{F}]
\right]
\end{equation}

where $d=\frac{1}{2}+\frac{1}{n-3}$ and $_{2}F_{1}$ is the usual
hypergeometric function.\\

If $t=t_{s}(r)$ stands for the time of collapse of a shell of
radius $R$ giving rise to a shell focusing singularity at $r$,\\
\begin{equation}
t_{s}(r)-t_{i}=\frac{2}{n-1}~r^{\frac{n-1}{2}}\frac{F^{\frac{a(n-1)-1}{2}}}{f^{\frac{b(n-1)}{2}}}
~_{2}F_{1}[\frac{1}{2},d,d+1,-r^{n-3}\frac{F^{a(n-3)-1}}{f^{b(n-3)-1}}]
\end{equation}

So the time of central  singularity is given by
\begin{equation}
\hspace{-3cm}t_{0}~=
\begin{array}{c}
\\
lim~~~t_{s}(r)\\
\hspace{-1cm}~r\rightarrow 0\\
\end{array}
=t_{i}+\frac{2F_{0}^{b}}{(n-1)f_{0}^{\frac{b(n-1)}{2}}}\\
\end{equation}

Here the power series expansions (12) and (13) have been used.\\

If $t=t_{ah}(r)$ stands for the time of formation of apparent
horizon at coordinate distance `$r$' then area radius is
restricted by [3,4]
\begin{equation}
R(t_{ah}(r),r)=[F(r)]^{\frac{1}{n-3}}
\end{equation}

and we have from (20)
\begin{equation}
t_{ah}(r)-t_{i}=\frac{2}{n-1}~r^{\frac{n-1}{2}}\frac{F^{\frac{a(n-1)-1}{2}}}{f^{\frac{b(n-1)}{2}}}
~_{2}F_{1}[\frac{1}{2},d,d+1,-r^{n-3}\frac{F^{a(n-3)-1}}{f^{b(n-3)-1}}]-\frac{2F^{\frac{1}{n-3}}}{n-1}~
_{2}F_{1}[\frac{1}{2},d,d+1,-f]
\end{equation}

Therefore, using the series expansions (12) and (13) the time
difference between central shell focusing singularity and the
apparent horizon at coordinate distance $r$, (close to $r=0$) is
given by

\begin{eqnarray*}
t_{ah}(r)-t_{0}=-\frac{d}{(n-1)(d+1)}\frac{F_{0}^{\frac{(3n-7)a-3}{2}}}{f_{0}^{\frac{(3n-7)b-2}{2}}}~
r^{\frac{1}{b}}-\frac{2F_{0}^{\frac{1}{n-3}}}{n-1}~r^{\frac{n-1}{n-3}}+
\frac{2bF_{0}^{b}}{(n-1)f_{0}^{\frac{b(n-1)}{2}}}\left[\frac{F_{2}}{F_{0}}+(b-1)\frac{F_{1}^{2}}{F_{0}^{2}}\right.
\end{eqnarray*}
\begin{equation}
\left.-\frac{(n-1)}{2}\frac{f_{2}}{f_{0}}+\frac{(n-1)}{8}\{b(n-1)+2\}\frac{f_{1}^{2}}{f_{0}^{2}}
\right]~r^{2}+O(r^{3})
\end{equation}

A little analysis of the equation (25) shows that when $b>1/2$ we
have $1/b<2$, so that there is only a black hole and no naked
singularity. The magnitude of $\frac{n-1}{n-3}$ is 3 at $n=4$ and
decreases with the increasing number of dimensions. For $b<1/2$
and $n<5$ there may occur either a black hole or a naked
singularity depending on the sign of the coefficient of $r^{2}$ in
equation (25). Again for the number of dimensions greater than
five, i.e., for six dimensions or more, $\frac{n-1}{n-3}<2$, so
that we get only a black hole. This conclusion is exactly
identical with that obtained in our previous paper [2].\\

{\bf Marginally bound case ($f=0$):}\\

Here our initial choice is $R=(r+r_{0})F^{a}$, $r_{0}$ and $a$
being positive constants. The density and shear are given by
\begin{equation}
\rho(t,r)=\frac{(n-2)F'(r)}{2R^{n-2}R'}~,~~~\sigma(t,r)=\sqrt{\frac{(n-2)}{8(n-1)}}~\left(
\frac{\dot{R}'}{R'}-\frac{\dot{R}}{R} \right)
\end{equation}

and the evolution equation simplifies to
\begin{equation}
\dot{R}^{2}=\frac{F(r)}{R^{n-3}}
\end{equation}

Suppose we perform linear transformations on $F$ and the area
radius $R$ in the following manner:
\begin{equation}
F\rightarrow p~F~,~~~~R\rightarrow q~R
\end{equation}

where $p,~q$ are constants. The invariance of $\rho,~\sigma$ and
the evolution equation (both initially and at any time instant)
under the above transformations yield
\begin{equation}
a=\frac{1}{n-1}~\text{and}~~p=q^{n-1}
\end{equation}

On the initial hypersurface, the energy density and the shear
scalar are given by
\begin{equation}
\rho_{i}=\frac{(n-2)}{2}\frac{\frac{rF'}{F}}{(r+r_{0})^{n-1}
\left(\frac{r}{r+r_{0}}+\frac{1}{n-1}\frac{rF'}{F} \right) }
\end{equation}
and
\begin{equation}
\sigma_{i}^{2}=\frac{(n-2)}{8}~\frac{(n-1)^{3}~r^{2}}{(r+r_{0})^{n+1}
\left[\frac{(n-1)r}{r+r_{0}}+\frac{rF'}{r} \right]^{2} }
\end{equation}

Here it is clear from the above that $r_{0}\ne 0$, because
otherwise there will be a permanent singularity of infinite energy
density and shear at the centre. In this case the solution of the
evolution equation has the simple form
\begin{equation}
t-t_{i}=\frac{2}{(n-1)}\left[(r+r_{0})^{\frac{n-1}{2}}-\frac{R^{\frac{n-1}{2}}}{\sqrt{F(r)}}
\right]
\end{equation}
So proceeding as before
\begin{equation}
t_{ah}(r)-t_{0}=r_{0}^{\frac{n-3}{2}}~r-\frac{2}{n-1}~F_{0}^{\frac{1}{n-3}}r^{\frac{n-1}{n-3}}~,
\end{equation}
which shows that $t_{ah}(r)>t_{0}$ for small $r$ (i.e., close to
singularity). Hence we always have naked singularity in any
dimension. Finally, it is to be noted that the parameter $r_{0}$
plays a crucial role in characterizing the end state of
collapse.\\

In conclusion we must mention that a significant point in our
paper is that we have avoided choosing $R(t_{i},r)=r$ as is done
in all the previous papers without perhaps any exception. Since
$R(t_{i},r)$ can be chosen any arbitrary function of $r$ without
any loss of generality we have preferred to select a different
function of $r$ for $R(t_{i},r)$, which allows us to perform a
different set of scalings and consequently draw different
conclusions in some situations. For example, the sign of the
quantity $(t_{ah}(r)-t_{0})$ in our present work depends upon an
additional parameter $b$, the suitable choice of which may yield
in certain cases different conclusions regarding the occurrence
of the naked singularity or black hole. However, the result we
obtained in one of our papers that there is no naked singularity
in the space-time having more than five dimensions, still holds
showing the possibility of this result being valid in general.\\

It is to be noted that in our paper even if the quantity
$(t_{ah}(r)-t_{0})$ is not invariant, its sign, however, will
remain unchanged under the scaling chosen in our paper, as one can
easily verify from the equation (25). So the conclusion about the
end result of the collapse (naked singularity or black hole) still
remains valid even after the scaling the parameters. Since it
depends on the sign and not the magnitude of the quantity
$(t_{ah}(r)-t_{0})$.\\

It is further to be stresses that for the parameter $b$ under
certain restriction $(b>1/2)$ the end state of the collapse always
leads to black hole irrespective of any number of dimensions of
our space-time. So finally we may conclude that with a suitable
choice of the initial area radius $R$ as a function of the
comoving coordinate $r$ we can ensure the validity
of the so-called Cosmic Censorship Conjecture.\\

{\bf Acknowledgement:}\\

One of the authors (SC) is thankful to CSIR, Govt. of India for
providing a research project No. 25(0141)/05/EMR-II.\\

{\bf References:}\\
\\
$[1]$ U. Debnath, S. Chakraborty and N. Dadhich, {\it Int. J. Mod.
Phys. D}  (accepted) 2005; {\it gr-qc}/0410031.\\
$[2]$ A. Banerjee, U. Debnath and S. Chakraborty, {\it Int. J.
Mod. Phys. D} {\bf 12} 1255 (2003).\\
$[3]$  S. G. Ghosh and N. Dadhich, {\it Phys. Rev. D} {\bf 64}
047501 (2001); S. G. Ghosh and A. Beesham, {\it Phys. Rev. D} {\bf 64} 124005 (2001).\\
$[4]$ U. Debnath and S. Chakraborty, {\it Gen. Rel. Grav.} {\bf
36} 1243 (2004).\\
$[5]$ F. C. Mena, B. C. Nolan and R. Tavakol, {\it gr-qc}/0405041.\\

\end{document}